\documentclass[letterpaper]{JHEP3}
\usepackage{graphicx}
\usepackage{amsmath}
\usepackage{amsfonts}
\usepackage{amssymb}
\usepackage{cite}

\newcommand{\eq}[1]{Eq.~(\ref{eq:#1})}
\renewcommand{\nc}{\newcommand}
\nc{\beq}{\begin{equation}}
\nc{\eeq}{\end{equation}}
\nc{\barray}{\begin{eqnarray}}
\nc{\earray}{\end{eqnarray}}
\nc{\barrayn}{\begin{eqnarray\star}}
\nc{\earrayn}{\end{eqnarray\star}}
\nc{\bcenter}{\begin{center}}
\nc{\ecenter}{\end{center}}
\nc{\ket}[1]{| #1 \rangle}
\nc{\bra}[1]{\langle #1 |}
\nc{\0}{\ket{0}}
\nc{\mc}{\mathcal}
\nc{\etal}{\emph{et al}}
\nc{\GeV}{\textnormal{GeV}}
\nc{\er}[1]{(\ref{eq:#1})}
\nc{\onehalf}{\frac{1}{2}}
\nc{\partialbar}{\bar{\partial}}
\nc{\psit}{\widetilde{\psi}}
\nc{\Tr}{\textnormal{Tr}}
\nc{\tc}{\tilde c}
\nc{\tk}{\tilde K}
\nc{\tv}{\tilde V}
\nc{\CN}{\mathcal{N}}

\nc{\Imtau}{\textnormal{Im}\,\tau}
\newcommand{\N}{\mathcal{N}}
\renewcommand{\t}{\tilde}
\newcommand{\del}{\partial}
\newcommand{\Del}{\nabla}
\newcommand{\comment}[1]{}

\title{The Universal K\"ahler Modulus in Warped Compactifications}

\author{Andrew R.~Frey\\ Department of Physics, McGill University,\\
 Montr\'eal, QC H3A 2T8 Canada} 
\author{Gonzalo Torroba\\ NHETC and
Department of Physics and Astronomy\\  Rutgers University. Piscataway,
NJ 08854, USA\\ and\\ 
Kavli Institute for Theoretical Physics,\\ 
University of California, Santa Barbara CA 93106, USA} 
\author{Bret
Underwood\\ Department of Physics, McGill University,\\ Montr\'eal, QC
H3A 2T8 Canada}
\author{Michael R.~Douglas\\ 
Simons Center for Geometry and Physics,\\ Stony Brook NY 11790, USA\\ and\\ 
NHETC and Department of Physics and
Astronomy\\  Rutgers University. Piscataway, NJ 08854, USA\\ 
and\\ I.H.E.S., Le Bois-Marie, Bures-sur-Yvette, 91440 France}

\abstract{ We construct the effective theory of the universal K\"ahler
modulus in warped compactifications using the Hamiltonian formulation of 
general relativity. The spacetime dependent 10d solution 
is constructed at the linear level for both the volume modulus
and its axionic partner, and nontrivial cancellations of warping effects 
are found in the dimensional reduction. Our main result is that the K\"ahler 
potential is not corrected by warping, up to an overall shift in the 
background value of the volume modulus. We extend the analysis
beyond the linearized approximation by computing the fully backreacted 
10d metric corresponding to a finite volume modulus fluctuation. 
Also, we discuss the behavior of the modulus in strongly warped regions
and show that there are no mixings with light Kaluza-Klein modes. 
These results are important for the phenomenology and cosmology of flux
compactifications.
}

\preprint{RUNHETC-2008-19\\NSF-KITP-08-133}  

\begin{document}

\section{Introduction}\label{sec:intro}

String backgrounds contain various light fields, such as metric zero
modes and spacetime tensors.  Determining their 4d dynamics,
\textit{i.e.}, dimensional reduction, is essentially a two-step
procedure. It requires first finding the correct 10d (in the case of
string theory) fluctuation corresponding to the 4d field. Then the 4d
action is computed by substituting this fluctuation ansatz into the
10d action.  In some  cases, the first step is simple, which can lead
to confusion in more complicated backgrounds.  In this paper, we
continue along the lines of \cite{dt, stud, dst} to advance the proper
treatment of dimensional reduction in conformally Calabi-Yau warped
compactifications of type IIB string theory 
\cite{hep-th/9605053,hep-th/9908088,hep-th/0004103,gkp}.

The moduli space by dimensional reduction 
is well understood for Calabi-Yau (CY) compactifications, mainly
because $\N=2$ supersymmetry determines the action in terms of a
single prepotential and the zero modes are associated to harmonic forms
of the CY \cite{Candelas:1990pi}.  However, Kaluza-Klein reduction in flux
compactifications with $\N=1$ supersymmetry is much less
understood. From a field theory point of view, finding the K\"ahler
potential is a complicated task because it is no longer protected by
holomorphicity. From a geometrical point of view, breaking $\N=2
\to\N=1$ corresponds to having a warped background, with the warp
factor sourced by branes, orientifold planes, and supergravity flux. 
Dimensionally reducing on these backgrounds is subtle, since the profiles
of the zero and higher Kaluza-Klein (KK) modes are nontrivial
\cite{hep-th/0201029,hep-th/0308156,da1,buchel,da2,gm}.

Understanding the dynamics in these cases is an important task,
because flux compactifications have many of the necessary ingredients
to produce realistic models of phenomenology and cosmology
\cite{hep-th/9605053,hep-th/9908088,hep-th/0004103,gkp}.  
Understanding that phenomenology therefore requires a knowledge
of the proper dimensional reduction.
For example, the 10d wavefunction controls the interactions of supergravity
moduli with brane fields, which may represent the Standard Model, as in
\cite{hep-th/0312232,hep-th/0409098}.  Additionally, supergravity KK modes
are dark matter candidates in some models; understanding their full
10d structure is important in determining whether their annihilation and
decay rates are sufficiently slow 
\cite{hep-th/0507257,hep-th/0602136,arxiv:0710.1299,Harling:2008px,arxiv:0802.2958}.

In this paper, we study dimensional reduction in
the compactifications of \cite{hep-th/9605053,hep-th/9908088,hep-th/0004103,gkp}.  
These 
are in some senses simpler than other flux
compactifications, both because the internal space is conformally CY
and because they have a well-defined supergravity limit.  Progress in
this direction has been made in \cite{gm,stud}, especially with
respect to volume-preserving fluctuations of the internal space. Recently,
in \cite{dt}, a formalism for computing kinetic terms in general
warped backgrounds was developed which makes the physical interpretation
of the computations manifest.  Since this formalism does not rely on
supersymmetry, it applies to conformally CY flux compactifications with
flux that breaks supersymmetry as well (see \cite{Burgess:2006mn} for some
discussion of supersymmetry breaking in these backgrounds).

Our goal in the present work is to elucidate the dynamics of the
universal K\"ahler modulus, applying the Hamiltonian-based method developed
in \cite{dt} (see also \cite{ku1,gm, ku2} for progress towards a 10d 
description). This mode arises in any string background with a
geometric interpretation, but its kinetic term has not yet been fully
understood in the case of general warping. A particularly important question 
is how warping 
effects correct the kinetic terms and
K\"ahler potential (for ${\mathcal N}=1$ theories). We will find that the K\"ahler 
potential is in fact 
not corrected by warping, up to an additive shift in the background value 
of the modulus.
This is a rather surprising outcome, because the 10d solution constructed 
from the Hamiltonian 
method is quite different from the unwarped fluctuation.  However, the 
needed shift in the modulus would affect nonperturbative superpotentials or
higher-derivative corrections that break the no-scale structure of the
classical background \cite{kklt,hep-th/0204254,hep-th/0408054,hep-th/0502058}.

Most of the methods developed so far apply to moduli dynamics in the
linearized approximation,  namely when the fluctuations around the
vacuum expectation values are  infinitesimal. This is  certainly
enough if one is interested in the K\"ahler potential and mass
spectrum of the theory.  However, understanding other effects,
particularly in cosmology, beyond the 4d effective field theory
requires going beyond the linearized level. For  this reason, we also
extend our approach to the case of finite  spacetime dependent
fluctuations of the volume modulus. This not only  should serve to
eliminate remaining confusion about the relation between the 10d and
4d theories, but it is also a significant first step in developing
cosmological solutions of  compactified 10d supergravity.  Such
solutions would demonstrate what signatures higher-dimensional or
string physics could be generated,  for example, by inflation in
string theory \cite{hep-th/0508139}.

Throughout, we restrict to conformally CY flux compactifications, but
our  method could be applied to more general $\N=1$ and
nonsupersymmetric  backgrounds as well.

\subsection{Beyond the Calabi-Yau case}\label{subsec:beyond}

Before starting our analysis, it is instructive to review the simpler
case of a Calabi-Yau compactification without warping. We follow the
discussion of  \cite{gkp} for IIB CY compactifications. The universal
volume modulus corresponds to a simple rescaling \beq \tilde g_{ij}
\to e^{2u}\,\tilde g_{ij} \eeq of the interal CY metric $\tilde
g_{ij}$. The time-dependent metric fluctuation is, at linear order,
\begin{equation}\label{eq:cy1}
ds^2=e^{-6u(x)}\,\eta_{\mu\nu}\,dx^\mu dx^\nu+ e^{2u(x)}\,\tilde
g_{ij}(y)\,dy^i dy^j\ ,
\end{equation}
where the 4d Weyl factor $e^{-6u(x)}$ is needed to decouple the
modulus  from the graviton. This 4d rescaling defines the 4d Einstein
frame and gives the Einstein-Hilbert action for the metric in 4d.  The
Einstein equations then reduce to the desired $\Box u=0$ for the
modulus. The 4-form field contributes an axion \beq C_4=
\frac{1}{2}a(x)\,\t J \wedge \t J + \cdots\ \eeq ($\t J$ is the fixed
K\"ahler form associated with the fixed CY metric  $\t g_{ij}$), which
pairs with the volume modulus into the complex field $\rho=a+i
e^{4u}$. Performing the dimensional reduction, one finds
\begin{equation}
K=-3\,\log\left(-i(\rho - \bar \rho)\right)\ .
\end{equation}

Backreaction from fluxes and branes (of the BPS type discussed in
\cite{gkp}) introduces warping to the background,
\begin{equation}\label{eq:intro-warped}
ds^2=e^{2A(y)}\,\eta_{\mu\nu}\,dx^\mu dx^\nu+e^{-2A(y)}\,\tilde
g_{ij}\,dy^i dy^j\,.
\end{equation}
One could then try different ways of identifying the universal volume
modulus. The simplest possibility would be to consider the same
dependence as in (\ref{eq:cy1}), even in the presence of warping
\cite{dg}:
\begin{equation}\label{eq:simple}
ds^2=e^{2A(y)}\,e^{-6u(x)}\,\eta_{\mu\nu}\,dx^\mu
dx^\nu+e^{-2A(y)}\,e^{2u(x)}\,\tilde g_{ij}(y)\,dy^i dy^j\,.
\end{equation}
This proposal does not work for a couple of reasons. Under a
spacetime-independent rescaling $\tilde g_{ij} \to e^{2u} \tilde
g_{ij}$, the warp factor acquires a dependence on $u$  \beq
e^{-2A}\,\to\,e^{-2u}\,e^{-2A}\eeq   in such a way that the full
internal metric $e^{-2A} \tilde g_{ij}$ is actually invariant under
the rescaling. Therefore, the simple rescaling of the CY metric
becomes a gauge redundancy which may be set to zero by a 4d Weyl
transformation. At a more technical level, \eq{simple} cannot solve
the 10d Einstein equations, so it does not give a consistent
time-dependent fluctuation.

Another possibility is suggested by the fact that the warp factor is
only determined up to an overall shift, \beq
e^{-4A(y)}\,\to\,e^{-4A(y)}+c\ .\eeq The volume of the compact space
scales as $V \sim c^{3/2}$, so it would be natural to identify this
flat direction as the warped version of the universal volume
modulus. One could then promote $c$ to a spacetime field $c(x)$ by
considering the metric fluctuation \cite{ku1,ku2,gm}
\begin{equation}
ds^2=\left[c(x)+e^{-4A_0} \right]^{-1/2}\,\eta_{\mu\nu}\,dx^\mu
dx^\nu+\left[c(x)+e^{-4A_0} \right]^{1/2}\,\tilde g_{ij}\,dy^i dy^j
\end{equation}
and performing the dimensional reduction.  However, this proposal does
not solve the linearized equations of  motion\footnote{Except for
special choices of $c(x)$ which appear to lead to instabilities \cite{ku1,ku2}.}  
either; additional
components of the metric are required to satisfy all the components of
the 10d Einstein equation \cite{gm}.  Dimensional reduction on
backgrounds for which the 10d equations of motion are not satisfied in
general does not lead to good low energy effective theories, and can
result in ambiguities, as noticed in previous studies
\cite{da1,buchel,da2,ku1,ku2}.


Summarizing, the dynamics of the universal K\"ahler modulus are not
understood beyond the CY case, and a more systematic approach is
needed. In this paper we will use the method proposed in \cite{dt} to
find the wavefunction for the volume modulus and its axionic partner
in the presence of warping. This approach can also be extended to more
general $\N=1$ or nonsupersymmetric backgrounds.

\section{Review of the Hamiltonian approach}\label{sec:review}

The main obstacle in computing the 4d action is the appearence of
``compensating'' fields \cite{gm, gray}.  These arise in any system
with gauge redundancies and time-dependent fields. In a Lagrangian
formulation their role is not manifest. If they are not taken into
account properly, the low energy effective action is not invariant
under 6d diffeomorphisms, making the description inconsistent.  In
\cite{dt} it was shown that a simple way of deriving the correct gauge
invariant action is in the Hamiltonian framework. The compensators are
then identified as Lagrange multipliers, and their dynamical role
becomes manifest. For completeness, in this section we summarize the
results of \cite{dt}.

\subsection{Gauge invariant fluctuations}

Consider a warped 10d background preserving 4d maximal symmetry,
\begin{equation}\label{eq:warped-product}
ds^2=e^{2A(y;\,u)}\,\hat g_{\mu \nu}(x) dx^\mu dx^\nu+ g_{ij}(y;u)
dy^i dy^j\,,
\end{equation}
which depends on metric zero modes $u^I$ (which do not mix with the 4d
metric at linear order).  The kinetic terms for $u^I$ are obtained by
promoting the modes to spacetime dependent fields $u^I(x)$.  However,
the new metric (\ref{eq:warped-product}) with spacetime dependent
$u^I(x)$ is generically (for non-trivial warp factor) no longer a
solution of the 10d Einstein equations.  In particular, the mixed
component of the Ricci tensor $R_{\mu i}$ acquires a term proportional
to $\partial_\mu u^I$ and becomes nonzero. Ans\"atze of this form are
therefore not viable starting points for KK dimensional reductions.

This problem is solved at linear order in velocities by including
compensators $\eta_{Ij}$,
\begin{equation}\label{eq:warped-comp}
ds^2=e^{2A(y;\,u)}\,\hat g_{\mu \nu}(x) dx^\mu dx^\nu+2\,\eta_{Ij}(y)
\partial_\mu u^I \,dx^\mu dy^j+ g_{ij}(y;u) dy^i dy^j\,.
\end{equation}
The spacetime dependent metric fluctuation (\ref{eq:warped-comp}) and
the  4d kinetic term are then obtained by solving the 10d Hamiltonian
equations  of the warped background. We refer the reader to \cite{adm}
for the  formulation of general relativity in canonical variables.

The formalism is based in computing the canonical momentum $\pi^{MN}$,
which may be seen to be equal to
\begin{equation}\label{eq:defpi}
(g^{tt}\,h)^{-1/2}\,\pi_{MN}=  \left( \dot h_{MN}-D_M \eta_N - D_N
\eta_M\right)-h_{MN}\,h^{PQ}\, \left( \dot h_{PQ}-D_P \eta_Q - D_Q
\eta_P\right)
\end{equation}
where $h_{MN}$ is the 9d space-like metric with components
$(g_{\mu\nu},\,g_{ij})$ for $\mu,\nu \neq 0$. $D_M$ is the covariant
derivative constructed from $h_{MN}$, and $\eta^N= \dot
u^I\,\eta^N_I$.  Then the Hamiltonian density becomes
\begin{equation}\label{eq:HG}
\mathcal H_G= \sqrt{-g_D} \left( -R^{(9)} +h^{-1} \pi^{MN} \pi_{MN}-
\frac{1}{8} h^{-1} \pi^2\right)-2 h^{1/2} \eta_N \nabla_M (h^{-1/2}
\pi^{MN})
\end{equation}
Notice that the compensating vectors $\eta^N= \eta^N_I\,\dot u^I$ only
appear as Lagrange multipliers, enforcing the constraints
\begin{equation}\label{eq:pi-constr}
D_N \left(h^{-1/2}\pi^{NM}\right)=0\,.
\end{equation}
After satisfying this, one can choose the gauge $\eta^N=0$, as usual
in constrained Hamiltonian systems. Therefore, in the Hamiltonian
framework, their dynamical role is manifest.

Notice that the time variation $\dot h_{MN}$ always appears combined
with the Lagrange multipliers, as in (\ref{eq:defpi}). For this
reason,  it is convenient to introduce the new metric fluctuation
\begin{equation}\label{eq:physh}
\delta_I h^{MN}:= \frac{\partial h^{MN}}{\partial u^I}-D^M \eta^N_I
-D^N \eta^M_I\;\,.
\end{equation}
Similarly, from the canonical momentum we define the variation
\begin{equation}\label{eq:pi-0m}
\dot u^I\, \delta_I \pi^{MN}:=2(g_{tt})^{1/2}\,\left(h^{-1/2}
\pi^{MN}\right)=\dot u^I\,\left(\delta_I h^{MN}-h^{MN}\,\delta_I h
\right)\,.
\end{equation}
The Hamiltonian constraint now becomes
\begin{equation}\label{eq:Dconstr}
D_N\left[(g^{tt})^{1/2}\,\delta_I \pi^{MN} \right]=0\,,
\end{equation}
which implies that $\delta_I \pi^{MN}$ is orthogonal to gauge
transformations.  The kinetic term extracted  from the Hamiltonian
reads
\begin{equation}\label{eq:generalK}
H_{kin}=\frac{1}{4}\,\dot u^I\,\dot u^J\, \left(\int d^{D-1} x
\sqrt{-g_{D}}\,g^{tt}\,\left[\delta_I \pi_{MN} \,\delta_J
\pi^{MN}-\frac{1}{D-2} \,\delta_I \pi \,\delta_J \pi \right] \right)\,,
\end{equation}
with $D=10$ in the present case. The kinetic term is entirely
determined by $\delta_I \pi^{MN}$, which is then interpreted as the
10d  gauge invariant metric fluctuation corresponding to the zero mode
$u^I(x)$.

\subsection{Kinetic terms}

Performing the dimensional reduction starting from
(\ref{eq:generalK}), the  constraint along the 4d directions sets
$\delta_I \pi^{\mu \nu}=0$, which is equivalent to
\begin{equation}\label{eq:deltaA}
\delta_I e^{2A}=-\frac{1}{2}\,e^{2A}\,g^{kl}\,\delta_I g_{kl}\,.
\end{equation}
Then the warp factor variation may be eliminated from $\delta
\pi_{ij}$ yielding,
\begin{equation}\label{eq:delta-pi}
\delta_I \pi_{ij}=\delta_I g_{ij}+\frac{1}{2} g_{ij}\,g^{kl}\,
\delta_I g_{kl}\,.
\end{equation}
The constraint along the internal directions,
\begin{equation}\label{eq:constr-0m}
D^N\big((g^{tt})^{1/2}\delta_I \pi_{Nj}\big))=0\,,
\end{equation}
implies that the physical fluctuation $\delta_I \pi_{ij}$ is in
harmonic gauge with respect to the full 10d warped metric.

With these results, the field space metric becomes \cite{dt}
\begin{equation}\label{eq:hamilton-G}
G_{IJ}(u)=\frac{1}{4} \int d^{6}y \sqrt{g_{6}}\,e^{2A}\,
\left(\delta_I \pi_{ij}\,\delta_J \pi^{ij} - \frac{1}{8}
\,g^{ij}\,\delta_I \pi_{ij}\,g^{kl}\, \delta_J \pi_{kl} \right)\,.
\end{equation}
The metric $G_{IJ}$ is given as an inner product (depending explicitly
on the warp factor)  between tangent vectors $\delta_I \pi_{ij}$ and
$\delta_J \pi_{ij}$. The condition  (\ref{eq:constr-0m}) implies that
the physical variation is orthogonal to gauge  transformations. An
equivalent statement is that the constraint equation minimizes  the
inner product over each gauge orbit. This is exactly what happens in
the simpler  Yang-Mills case, where the canonical momentum is the
electric field, the constraint  is Gauss's law, and the kinetic term
is proportional to the electric energy.

This method applies to general warped compactifications preserving 4d
maximal symmetry. No assumptions about the internal metric $g_{ij}(y)$
in \eq{warped-product} are required. The application of these results
to the particular case of a conformal Calabi-Yau metric
(\ref{eq:intro-warped}) is discussed in \cite{dt} and will be
summarized in section \ref{subsec:orthog}.

\section{Finding the universal volume modulus}\label{sec:find}

Our aim is to find the 10d solution describing a finite spacetime
dependent  fluctuation of the volume modulus. Now, as explained in
section  \ref{subsec:beyond},  the first problem one faces is that of
defining the volume modulus in warped  backgrounds. We address this
issue by finding the modulus in the case of an  infinitesimal
fluctuation, and then showing how to integrate it to a finite
variation in section \ref{sec:ppwave}.

Before proceeding, we should clarify the type of expansion being
performed.  One starts from a warped background of the general form
(\ref{eq:warped-product}),  where $\hat g_{\mu\nu}$ is a maximally
symmetric 4d metric. Then, a given  modulus $u$ is allowed to have a
nontrivial spacetime dependence, acquiring  a nonzero velocity $\dot
u$ and energy $g^{tt}\,(\dot u)^2$. The energy  sources the Ricci
tensor, with the result that maximal symmetry is lost;  for instance,
for a massless excitation we would have a pp-wave spacetime.  The
important point is that backreaction is proportional to the energy,
and hence is quadratic in $\dot u$. The linearized expansion we
consider  here then means working at first order in moduli velocities,
so that the  4d metric can still be approximated by a maximally
symmetric space. In  this limit, the metric fluctuations $\dot
h_{MN}=\dot u^I\,\partial_I h_{MN}$  amount to a small perturbation
around the background solution $h_{MN}$  even if $\partial_I h_{MN}$
is not necessarily small. This is enough for  the purposes of finding the
K\"ahler potential.

We apply the Hamiltonian approach to find the linearized 10d
wavefunctions of the universal volume modulus (in this section) and
its axionic partner (in section \ref{sec:axion}). These results will
be  used in section \ref{sec:4d} to compute the K\"ahler potential.
Finally, in section \ref{sec:ppwave} we extend our results beyond the
linear  approximation, finding the backreaction produced by a finite
volume  modulus fluctuation. We restrict to type IIB with BPS fluxes
and  branes \cite{hep-th/9605053,hep-th/9908088,hep-th/0004103,gkp}, so that the internal manifold is conformally
equivalent to a Calabi-Yau:
\begin{equation}\label{eq:warped-product2}
ds^2=e^{2A_0(y)}\,\hat \eta_{\mu\nu}\, dx^\mu dx^\nu+ e^{-2A_0(y)}
\tilde g_{ij}(y) dy^i dy^j
\end{equation}
(thus, we work in the orientifold limit with constant axio-dilaton as
well).  It would be interesting to apply our approach to general
$\mathcal N=1$ flux  compactifications.

\subsection{Ten dimensional wavefunction}\label{subsec:computation}

Consider an ansatz of the form (\ref{eq:warped-comp}),
\begin{equation}\label{eq:warped-c}
ds^2=e^{2A(y;\,c)+2 \Omega[c]}\,\left(\hat g_{\mu \nu}(x) dx^\mu
dx^\nu+2 \,\partial_j B\,\partial_\mu c \,dx^\mu dy^j\right)+
e^{-2A(y;\,c)}\,\tilde g_{ij}(y) dy^i dy^j\,,
\end{equation}
where $c(x)$ denotes the universal volume modulus.  As will be seen
momentarily, a compensating field proportional to a total derivative,
\begin{equation}\label{eq:comp-ansatz}
\eta_j(y)=e^{2A+ 2 \Omega}\,\partial_j B(y)\ ,
\end{equation}
solves the Hamiltonian constraints, so we have already made this
identification in the ansatz. The Weyl factor is defined to bring us
to $4$-dimensional Einstein frame,
\begin{equation}\label{eq:Omega-def}
e^{2\Omega(c)}=\frac{\int \,d^6y \,\sqrt{\tilde g_6}}{\int \,d^6y
\,\sqrt{\tilde g_6}\,e^{-4A(y;\,c)}}=\frac{V_{CY}}{V_W(c) }\,.
\end{equation}
Furthermore, the underlying CY metric is taken to be independent of
the  volume modulus because a rescaling $\tilde g_{ij} \to \lambda
\tilde g_{ij}$  amounts to a 4d Weyl transformation.

At the end of the section it will be argued that $c(x)$ is actually
orthogonal to the  other non-universal metric zero modes $u^I(x)$. It
is then consistent to set these to  zero in the present
discussion. Next we will show how the Hamiltonian approach determines
the 10d wavefunction (\ref{eq:warped-c}). 
The full computation is somewhat technical, so 
in section \ref{subsec:summary} we summarize the
results.

The first step is to compute the canonical momentum (\ref{eq:pi-0m})
associated to the ansatz \eq{warped-c}). These are found to be
\begin{eqnarray}
\delta_c \pi_{\mu\nu}&=& 2\,h_{\mu\nu}\,\left(4\, \frac{\partial
A}{\partial c} - 2\, \frac{\partial \Omega}{\partial c} + \nabla^i
\eta_i+ 2 \,\partial^i A \, \eta_i \right)\ \textnormal{and}
\nonumber\\  \delta_c\pi_{ij}&=& g_{ij} \left(4\, \frac{\partial
A}{\partial c} - 6 \,\frac{\partial \Omega}{\partial c}+2\,\nabla^i
\eta_i+6\, \partial^i A\, \eta_i \right) -\nabla_i \eta_j -\nabla_j
\eta_i\ ,
\end{eqnarray}
where $\eta_i$ is given in (\ref{eq:comp-ansatz}).

Since $\delta_c \pi_{\mu\nu}$ is proportional to $h_{\mu\nu}$, the
constraint $D^N \,\pi_{N\nu}=0$ sets $\partial^\mu\left(\delta_c
\pi_{\mu\nu}\right)=0$. This relation should be valid for arbitrary
$\partial_\mu c$, implying $\delta_c \pi_{\mu\nu}=0$, or
\begin{equation}\label{eq:constrB}
4\, \frac{\partial A}{\partial c} - 2\, \frac{\partial
\Omega}{\partial c} + e^{2\Omega+4A}\,\tilde\Del^{2}B=0\,,
\end{equation}
in terms of the derivative $\tilde \nabla_i$  and Laplacian compatible
with $\tilde g_{ij}$.

The constraint $D^N\,\pi_{Nj}=0$ requires a bit more of work.
Fortunately, we can use the computation of the Ricci tensor component
$R_{\mu i}$ in \cite{ku1, gm} for our purposes,  recalling the
relation \cite{adm} \beq R_{0 i}=-D^N(h^{-1/2}\,\pi_{Ni})\,.  \eeq 
(We
also need a diffeomorphism transformation to set $\eta_i=0$ and
$\eta_\mu=- e^{2A+2\Omega}\,\partial_\mu \dot c\,B$, which can always
be done for a compensator of the form (\ref{eq:comp-ansatz})). The constraint
then sets
\begin{equation}\label{eq:shift}
\partial_m \big( \partial_c e^{-4A(y;\,c)}\big)=0,
\end{equation}
which implies that the dependence of the warp factor on $c(x)$ is
given by an additive shift
\begin{equation}\label{eq:A-c}
e^{-4A(y;\,c)}=e^{-4A_0(y)}+ c(x)\,,
\end{equation}
where $e^{-4A_0(y)}$ denotes the solution associated to the metric
$\tilde g_{ij}$, which is independent of $c(x)$. A possible
multiplicative  factor is fixed using the integrated version of
(\ref{eq:constrB}).

This result has an intuitive interpretation. In conformally CY flux
compactifications, the background equations of motion only fix
$e^{-4A}$  up to a shift $e^{-4A} \to e^{-4A} + c$. It was noticed in
\cite{ku1, ku2, gm}  that a change in $c$, which is not a simple
metric rescaling, also changes the internal volume, leading to the
proposal that $c$ represents the time-independent universal volume
modulus. What we find here is that this shift is present in the full
time-dependent case too, although the full 10d metric fluctuation has
other components as well.

Finally, plugging (\ref{eq:Omega-def}) and (\ref{eq:A-c}) into
(\ref{eq:constrB}),  we obtain the differential equation that fixes
the compensating field (also observed in \cite{gm}),
\begin{equation}\label{eq:constr4d}
\tilde \nabla^2 B=-e^{-4A-2\Omega}\,\left(4\,\frac{\partial
A}{\partial c} -2 \frac{\partial \Omega}{\partial
c}\right)=-e^{-4A_0}+\frac{V_W^0}{V_{CY}}\,,
\end{equation}
where $V_W^0 = \int d^6y \sqrt{\tilde{g}_6} e^{-4A_0(y)}$ is the
background value of the warped volume.  This equation is consistent in
compact CY manifolds because the right hand side integrates to zero
(which is actually the condition which fixes the factor of
$e^{2\Omega}$ in (\ref{eq:comp-ansatz})). Therefore,  the 10d metric
solving the Hamiltonian constraints,
\begin{eqnarray}\label{eq:10dlinear}
ds^2_{10} &=& \left[e^{-4A_0(y)}+ c(x)\right]^{-1/2} e^{2\Omega[c(x)]}
	\,\left(\hat{g}_{\mu \nu}(x)\,dx^\mu dx^\nu +2 \,\partial_i
	B\,\partial_\mu c\,dy^i dx^\mu\right)+\nonumber \\
	&&+\left[e^{-4A_0(y)}+c(x)\right]^{1/2}
	\,\tilde{g}_{ij}(y)\,dy^i dy^j
\label{eq:App1FluctMetricConstraint}
\end{eqnarray}
gives a consistent spacetime dependent solution representing
infinitesimal  fluctuations of the universal volume modulus. The last
part of the 10d  fluctuation is in the 4-form potential, which is
proportional to $e^{4A}$.  Intuitively, the BPS-like condition of
\cite{gkp} sets  $C_4= e^{4\Omega}\,e^{4A(y;\,c)}\,d^4 x$, so the
4-form fluctuates along  with the volume modulus. More details are
given in section \ref{sec:ppwave}, where these results will be extended to
finite  fluctuations.

\subsection{Summary}\label{subsec:summary}

Briefly summarizing the main points of the previous computation, the
warped universal volume modulus is not associated to a simple trace
rescaling of the underlying CY metric, unlike in the unwarped case.
Rather, $\tilde g_{ij}(y)$ stays  fixed and the modulus corresponds to
an additive shift \beq e^{-4A(x,y)}=e^{-4A_0(y)}+c(x)\ , \eeq where
$e^{-4A_0(y)}$ is the background solution with respect to $\tilde
g_{ij}$. There is also a nonzero compensating field $\partial_i B$
determined by (\ref{eq:constr4d}).

A more physical way of stating this is by noticing that in the 4d
action the compensating field \emph{only} appears through the shift
\cite{dt}
\begin{equation}
\delta_c g_{MN}=\frac{\partial g_{MN}}{\partial c}-D_N
\left(e^{2A+2\Omega}\,\partial_M B \right)-D_M
\left(e^{2A+2\Omega}\,\partial_N B \right)\,,
\end{equation}
where $D_N$ is the covariant derivative with respect to the 9d spacelike metric (see section \ref{sec:review}).
The physical 10d fluctuation associated to $c(x)$ then becomes
\begin{equation}\label{eq:delta-c-g}
\delta_c g_{\mu\nu}= 2\,e^{2A+2\Omega}\,\eta_{\mu\nu}\left(\delta_c A
+ \frac{\partial \Omega}{\partial c}\right)\,,\; \delta_c
g_{ij}=-\,e^{-2A} \left(2 \,\delta_c A\,\tilde g_{ij}+\delta_c  \tilde
g_{ij}\right)\ ,
\end{equation}
where
\begin{equation}\label{eq:deltaA-g}
\delta_c A:=\frac{\partial A}{\partial
c}-e^{4A+2\Omega}\,\partial^{\tilde \imath}A\,\partial_i
B\,,\;\;\delta_c \tilde g_{ij}=\tilde \nabla_i\left[e^{4A+2\Omega}
\partial_j B\right]+\tilde \nabla_j\left[e^{4A+2\Omega} \partial_i
B\right]\,.
\end{equation}
The dependence of $\Omega$ and $A$ on $c(x)$ is given in
(\ref{eq:Omega-def}) and (\ref{eq:A-c}).  Strikingly, for non-trivial
warping the universal volume modulus has an internal metric
fluctuation $\delta_c g_{ij}$  which \emph{is not pure trace}. The
nontrivial dependence comes from the effect of the compensating
field. Stated in gauge invariant terms, this is required so that the
canonical momentum  $\delta_c \pi_{MN}$ built from $\delta_c g_{MN}$
is in harmonic gauge with respect to the warped 10d metric.

Notice that in the unwarped (or large volume)  limit the warp factor
becomes $e^{-4A} \approx c(x) := e^{4u(x)}$,  which in turn implies
$e^{2\Omega} = e^{-4u(x)}$.  The equation of motion for the
compensator (\ref{eq:constr4d}) becomes simply $\tilde\Del^2 B = 0$,
which is  solved by $B = 0$, so we regain the usual metric for the
universal volume modulus in the unwarped case (\ref{eq:cy1}).

\subsection{Orthogonality with other modes}\label{subsec:orthog}

The metric moduli arise as independent solutions to a Sturm-Lioville
problem.  Different zero modes should be orthogonal to each other, and
we may use this to understand how to define the universal volume
modulus from the original $h^{1,1}$ moduli.

The natural inner product is given by the Hamiltonian
(\ref{eq:generalK}). Consider two zero mode solutions, with canonical
momenta $\delta_I \pi_{MN}$ and $\delta_J \pi_{MN}$ respectively ($I
\neq J$). The orthogonality condition reads
\begin{equation}\label{eq:orthog}
G_{IJ}=\int d^{D-1} x\sqrt{-g_D}\,g^{tt}\,\big[\delta_I \pi_{MN}
\,\delta_J \pi^{MN}-\frac{1}{D-2} \,\delta_I \pi\,\delta_J \pi \big]
=0\,,
\end{equation}
where $D=10$ in our case.

We need to compute the inner product (\ref{eq:orthog}) between the
universal volume modulus and the nonuniversal metric fluctuations.
Recall that the canonical momentum associated to  such a fluctuation
is \cite{dt}
\begin{equation}\label{eq:bg2}
\delta_I \pi_{ij}=e^{-2A} \left(\delta_I \tilde g_{ij}-\frac{1}{2}
\tilde g_{ij}\, \delta_I \tilde g \right)\,,
\end{equation}
where
\begin{equation} \label{eq:bg3}
\delta_I \tilde g_{ij}=\frac{\partial \tilde g_{ij}}{\partial
u^I}-\tilde \nabla_i\left( e^{4A}\,B_{Ij}\big)-\tilde \nabla_j\big(
e^{4A} \,B_{Ii} \right)\,.
\end{equation}
Here $B_{Ij}$ is the compensating field required by the time-dependent
fluctuation $\partial \tilde g_{ij}/\partial u^I$. Unlike the case of
the universal modulus, the $B_{Ij}$ are not total derivatives; compare
with \eq{delta-c-g} and \eq{deltaA-g}.

Next, specialize to $I=c$, the universal volume modulus, and $J \neq
c$ a nonuniversal zero mode. Using orthogonality with respect to gauge
transformations and $\delta \pi_{\mu\nu}=0$, \beq G_{cJ}= \int d^{D-1}
x\sqrt{-g_D}\,g^{tt}\,\frac{\partial g^{ij}}{\partial c}\,\delta_J
\pi_{ij}\,.  \eeq  Recalling that $\partial g_{ij}/\partial
c=(1/2)\,e^{4A}\,g_{ij}$,  the orthogonality condition requires
\begin{equation}\label{eq:int-condition}
\int d^6y\,\sqrt{\tilde g_6}\,\tilde g^{ij}\,\delta_J \tilde
g_{ij}=0\,,
\end{equation}
which is solved by
\begin{equation}\label{eq:traceless}
\tilde g^{ij}\,\frac{\partial \tilde g_{ij}}{\partial u^J}=0\,.
\end{equation}
The compensating fields in (\ref{eq:bg3}) drop from
(\ref{eq:int-condition}), being total derivatives.

The nonuniversal K\"ahler moduli thus correspond to the $h^{1,1}-1$
\emph{traceless} combinations, and \eq{traceless} defines the  basis
of linearly independent metric zero modes orthogonal to the universal
volume modulus. It is interesting that we recover the known result
from  CY compactifications, although the universal mode is no longer a
pure trace  fluctuation of the internal metric. We should also point
out that  (\ref{eq:traceless}) is not a gauge condition: we can fix
completely the  diffeomorphism redundacies by setting the compensating
fields to zero,  but we would still need to impose
(\ref{eq:traceless}). Rather, it tells  us how to choose a particular
basis in the space of solutions to the  Sturm-Liouville problem of the
metric zero modes. This grants that there  are no kinetic mixings
between the volume modulus and the other zero modes.

\section{Axionic partner of the volume modulus}\label{sec:axion}

In the unwarped limit, the universal volume modulus gets complexified
with the axion coming from  \beq C_4= \frac{1}{2}a(x)\t J(y)\wedge\t
J(y)+\cdots\ .  \eeq   In this section, we construct the universal
axion in warped backgrounds. This will be the partner of the warped
volume modulus (\ref{eq:delta-c-g}). At the end of the section, the
$h^{1,1}-1$ nonuniversal axions will be shown to be orthogonal to the
universal axion, so they will be set to zero in the main part of the
analysis. This is the counterpart of what happens with the universal
volume modulus, as can be anticipated for supersymmetric
compactifications.

The Hamiltonian formulation for antisymmetric tensors is similar to
the  familiar $U(1)$ Maxwell case, where the canonical momentum is the
electric field,
\begin{equation}\label{eq:maxwell}
E^i=\frac{\partial \mathcal L}{\partial \dot
A_i}=g^{tt}\,g^{ij}\left(\partial_0 A_j - \partial_j A_0\right)\ ,
\end{equation}
and $A_0$ is a Lagrange multiplier enforcing Gauss's law $\nabla^i
E_i=0$. The shift of (\ref{eq:maxwell}) by $\partial_i A_0$ is the
analog of the metric fluctuation shift (\ref{eq:physh}) by the
compensating field.

The generalization to a $p$-form $C_p$ is as follows. $C_{0 i_2\ldots
i_p}$ plays the role of a Lagrange multiplier, and the canonical
momentum is given by the $p+1$-form
\begin{equation}\label{eq:E-def}
E:=\frac{1}{(p+1)!}\,F_{0 i_1 \ldots i_p}\,dx^0 \wedge dx^{i_1} \wedge
\ldots dx^{i_p}\,,
\end{equation}
where $i_1,\ldots, i_p$ are spacelike indices. If there are no
couplings  to external fields the constraint is
\begin{equation}\label{eq:E-constr}
d \left(\star_D\,E \right)=0\,.
\end{equation}
The Hamiltonian kinetic term is then
\begin{equation}\label{eq:E-kin}
\int dt\,H_{kin}= \int E \wedge \star_D\,E=\frac{1}{(p+1)!}\,\int d^D
x\, \sqrt{g_D}\, F_{0 i_1 \ldots i_p}\,F^{0 i_1 \ldots i_p}\,.
\end{equation}
This is gauge invariant due to (\ref{eq:E-constr}). The magnetic field
contributions $F_{i_1 \ldots i_{p+1}}$ appear in the potential energy.

\subsection{Axion fluctuation in a warped background with flux}

We now apply the previous approach to find the 10d universal axion in
the warped background (\ref{eq:warped-product2}) with three-form flux;
there are extra subtleties arising from self-duality and the unusual
gauge transformations of the 4-form potential.   We start by
generalizing the known form from unwarped compactifications, since the
wavefunction should reduce to that form in the unwarped limit.  We
also find that a constant axion yields a trivial field strength, even
in the presence of a fluctuating volume modulus, so the solution
respects the classical axion shift symmetry.  Also, recall that we are 
working in the limit of constant axio-dilaton.

Because the 4-form potential transforms under gauge transformations
associated with the 2-form potentials, there is a small subtlety in
determining 4-form fluctuations that are globally defined on the
compactification \cite{hep-th/0201029}.  We discuss the details in
Appendix \ref{sec:5formdef}; we will find that, in terms of globally
defined fluctuations, the 5-form and 3-form canonical momenta
(\ref{eq:E-def}) are
\begin{eqnarray}
\t E_5 &=& d \delta C_4 + \frac{ig_s}{2} (\delta A_2\wedge \bar{G}_3 -
\delta\bar{A}_2 \wedge G_3) \
\label{eq:CanonicalE5Def}\\ E_3 &=& d \delta A_2\ ,
\label{eq:CanonicalE3Def}
\end{eqnarray}
where $A_2 = C_2 - \tau B_2$. $\delta C_4$ and $\delta A_2$ denote the 
components of $C_4$ and $A_2$ which depend on the axion field; their 
explicit form will be given momentarily.  
The presence of the ``transgression terms" in
(\ref{eq:CanonicalE5Def}) reflects the fact that the canonical momenta 
are invariant under the gauge transformations
\begin{eqnarray}
\delta C_4 &\rightarrow & \delta C_4 + d \chi_3  + \frac{ig_s}{2}
(\bar{\zeta}_1\wedge G_3 - \zeta_1\wedge\bar{G}_3)\ ,  \nonumber\\
\delta A_2 &\rightarrow & \delta A_2+d\zeta_1\, .
\end{eqnarray}
We expect the axion to descend from the 4-form gauge potential $\delta
C_4$; however, we notice that there are two separate gauge
transformations associated with $\delta C_4$, one of which arises from
gauge transformations of $\delta A_2$.  From the Hamiltonian
perspective, gauge transformations are associated with corresponding
compensators, so we expect that there should be compensators for the
axion associated with \emph{both} $\delta C_4$ and $\delta A_2$.

We take the ansatz
\begin{equation}
\label{eq:axion1}
\delta C_4 = \frac{1}{2}a_0(x)\t J^2 +a_2(x)\wedge \t J -da_0\wedge
K_3  -da_2 \wedge K_1\ ,\ \  \delta A_2 = -da_0\wedge \Lambda_1
\end{equation}
(note that $\t J\wedge \t J=2\,\t\star_6\t J$).   Here, $a_0$ and
$a_2$ are spacetime 0- and 2-forms  respectively, while $K_{1,3}$ and
$\Lambda_1$ are forms on the internal manifold included as possible
compensators.  The canonical momenta
(\ref{eq:CanonicalE5Def}-\ref{eq:CanonicalE3Def}) are then
\begin{eqnarray}
\t E_5 &=& da_0\wedge \left(\t\star_6\t J+dK_3
-\frac{ig_s}{2}\Lambda_1\wedge \bar G_3
+\frac{ig_s}{2}\bar\Lambda_1\wedge G_3\right)+da_2\wedge \left(\t
J+dK_1\right) \\ E_3 &=& da_0 \wedge d\Lambda_1\, .
\end{eqnarray}
Notice that  $\t E_5$ vanishes trivially for a  constant axion $a_0$,
so the field space metric cannot depend on the axion, as expected from
the classical axion shift symmetry.  The 5-form canonical momentum $\t
E_5$ is self-dual, which reduces the 4d degrees of freedom to a single
scalar by requiring $da_2\propto \hat\star_4 da_0$.  At linear order,
the proportionality constant may depend only on expectation values of
moduli (at higher orders, it may also depend on  fluctuations of
moduli); we will see that the full wavefunction requires the choice
$da_2= e^{4\Omega}\hat\star_4 da_0$. 
In this work we only keep $a_0$ as an independent field,  multiplying
the kinetic term by $2$.\footnote{See \cite{moore} for a careful
treatment of the self-dual form.}

Imposing the constraint (\ref{eq:E-constr}) for the 5-form, we find
that
\begin{eqnarray}
\label{eq:constraints}
&& d\left[ e^{4A} \left(\t J + \t \star_6 \left(dK_3 -
\frac{ig_s}{2}\Lambda_1\wedge \bar G_3
+\frac{ig_s}{2}\bar\Lambda_1\wedge G_3\right)\right) \right] = 0 \\ &&
d \left[ e^{-4A} \left(\t \star_6 \t J + \t \star_6 dK_1\right)\right] =
-\frac{ig_s}{2} e^{-2\Omega} \left(d\Lambda_1\wedge \bar G_3-
d\bar\Lambda_1\wedge G_3\right)\, .\label{eq:constraints2}
\end{eqnarray}
These constraints are identical to the 10d equations of motion
$d(\star_{10}\t F_5) =(ig_s/2)G_3\wedge \bar G_3$ evaluated for legs
in the internal directions.  (The factor of $e^{-2\Omega}$ on  the
right-hand-side of \eq{constraints2} is related to the proportionality
factor in the 4d Poincar\'e duality between $a_0$ and $a_2$.)  In this
way, the Hamiltonian and Lagrangian approaches yield equivalent
results, and  $a_0$ corresponds to a massless 4d field.

For the volume modulus, the compensating field is  determined by a
single scalar function $\eta_i=e^{2A+2\Omega}\,\partial_i B$, and we
expect the same to occur  for the compensator in $\delta C_4$.  The
form of the compensator equation (\ref{eq:constraints})  then
motivates the following ansatz,
\begin{equation}
e^{4A} \left[\t J + \t \star_6 \left(dK_3 -
\frac{ig_s}{2}\Lambda_1\wedge  \bar
G_3+\frac{ig_s}{2}\bar\Lambda_1\wedge G_3\right)\right] =  e^{2\Omega}
\t J + e^{2\Omega} d\left(e^{4A} dK\right)
\label{eq:ansatz}
\end{equation}
in terms of a function $K(y)$. The factor of $e^{2\Omega}$ is fixed by
wedging (\ref{eq:ansatz}) with $\t \star_6 \t J$ and integrating over
the internal space.  In fact, this ansatz yields an appropriately
self-dual 5-form if we take $K_1=e^{4A}dK$, and the factor here
precisely fixes the proportionality in the relation between $a_0$ and
$a_2$.  \comment{This factor also leads to the correct unwarped limit
since with $e^{2\Omega} \sim e^{4A}\sim c^{-1}$ the terms proportional
to $\t J$ cancel,  and the equation admits the trivial solution
$K=0$.}  Replacing this ansatz in (\ref{eq:constraints2}), we obtain
the  compensator equation for $K(y)$,
\begin{equation} 
d\left(\t\star_6 [dA \wedge dK]\right) +\frac{1}{8}\,de^{-4A}\wedge\t J
\wedge \t J = -e^{-2\Omega} \frac{ig_s}{8}\left(d\Lambda_1\wedge \bar
G_3-d\bar\Lambda_1\wedge G_3 \right)\ .
\label{eq:c-ansatz}
\end{equation}

The second constraint, associated with the $A_2$ gauge transformation,
fixes the compensator $\Lambda_1$, \comment{
\beq\label{eq:g3constraint1} d\left(\star\delta G_3 - ig_s\delta
C_4\wedge G_3\right)=0\eeq or} \beq\label{eq:g3constraint2} d(\t\star_6
d\Lambda_1)=-4i\, e^{2\Omega}e^{4A}dA\wedge dK\wedge G_3\ .  \eeq
This follows from the $G_3$ equation of motion, primitivity of the 
3-form,\footnote{On an orientifold $T^6$ or
$T^2\times K3$, an additional term may appear in \eq{g3constraint2} if
the flux breaks supersymmetry, but it cancels out of the following
analysis.}
and the 4d Poincar\'e duality relation (which fixes the power
of $e^{2\Omega}$.

There is one other issue in this analysis.  Because there is a
background 5-form associated with the warp factor, the axion
fluctuations can appear in the Hamiltonian equation for $\dot
\pi_{MN}$ at  linear order,   through terms of the form $\delta\t
F_{MP_1\ldots P_4} \,\t F_N{}^{P_1 \ldots P_4}$. By examining the
allowed components, we can see that the only terms that contribute are
of the form
\begin{equation}\label{eq:f5einstein}
4\t F_\mu{}^{\nu\lambda\rho n}\delta \t F_{m\nu\lambda\rho n} +\t
F_m{}^{npqr}\delta \t F_{\mu npqr}\ .
\end{equation}
However, with the background 4-form potential proportional to the 4d
volume form, self-duality of the 5-form causes this contribution to
vanish for any fluctuations $\delta\t F$ with these components.

Summarizing, the gauge invariant wavefunction for the universal axion
in a warped background is given by the canonical momenta
\begin{eqnarray}
\label{eq:axion2}
\t E_5 &=& (1+\star_{10})\left[e^{2\Omega}da_0(x)\wedge  \t
\star_6\,\left(e^{-4A}\,\t J +4\, dA\wedge dK\right)\right] \\ E_3 &=&
da_0 \wedge d\Lambda_1\ ,
\end{eqnarray}
where $K,\Lambda_1$ satisfy the Gauss law constraints
(\ref{eq:c-ansatz},\ref{eq:g3constraint2}) respectively.
Heuristically, the warp factor dependence arises naturally from  $J
\wedge J = e^{-4A}\,\t J \wedge \t J$.

In the unwarped limit, we see that the compensators become gauge
trivial.  First, $K_1$ becomes exact. Similarly \eq{g3constraint2}
implies that $\Lambda_1$ is closed, so $\delta G_3=0$.   The residual
gauge freedom to make $\Lambda_1$ co-closed means that it must vanish
(because there are no harmonic 1-forms on a CY); this same gauge
transformation also forces $K_3$ to be closed, as required by
\eq{ansatz} since $e^{2\Omega} = e^{4A}= c^{-1}$.  Then it is simple
to gauge away the $K_1$ and $K_3$ compensators in  \eq{axion1}.  As
expected, we then recover the known axion wavefunction in a CY
background.  Also note that the compensators $\Lambda_1$ become
trivial when the  background 3-form flux vanishes, which we expect
because $\delta C_4$ has only one gauge transformation in that case.

\subsection{Orthogonality with nonuniversal axions}

We now consider the effect of the $h^{1,1}-1$ nonuniversal axions. The
story is similar to the above. For $\t \rho_r$ the independent $(1,1)$
forms in the 2nd cohomology ($\t J=\t\rho_1$),  the potential now
becomes
\begin{equation}\label{eq:manyaxions}
\delta C_4 = \sum_{r=1}^{h^{1,1}}\left[ a^r_0(x) \t \star_6 \t \rho_r +
a_2^r(x)\wedge \t \rho_r - da_0^r\wedge dK_{3,r} - da_2^r\wedge
dK_{1,r}\right]\,, \ \ \delta A_2 = -da_0^r\wedge \Lambda_{1,r}\, .
\end{equation}

Computing the canonical momentum, we obtain constraints analogous to
(\ref{eq:constraints},\ref{eq:constraints2}),
\comment{\begin{eqnarray}\label{eq:C-constr2} &&d\left[e^{4A}\left(\t
\rho_r+\t\star_6( dK_{3,\,r}- \frac{ig_s}{2}\Lambda_{1,r}\wedge \bar
G_3 +\frac{ig_s}{2}\bar\Lambda_{1,r}\wedge G_3)\right)\right]=0\\
&&d\left[e^{-4A}\left(\t\star_6\t \rho_r+\t\star_6\,
dK_{1,\,r}\right)\right]=0\,.
\end{eqnarray}}
which along with self-duality imply
\begin{equation}\label{eq:manyconstraint}
e^{4A}\left(\t \rho_r+\t\star_6 dK_{3,\,r}+
\frac{ig_s}{2}\Lambda_{1,r} \wedge \bar
G_3-\frac{ig_s}{2}\bar\Lambda_{1,r}\wedge G_3)\right)=
e^{2\Omega}M_r^s(u)\,\left(\t \rho_s+dK_{1,\,s}\right)\,,
\end{equation}
with $M(u)$ some function of the moduli $u^I$, which can be
diagonalized.  The constraint from the 2-form gauge transformation
is of a similar form as (\ref{eq:g3constraint2}), but with a more general
1-form $K_1 \neq e^{4A} dK$ on the right hand side,
because there are no harmonic 5-forms
on a CY.\footnote{Again, there are additional terms on $T^6$ or
$T^2\times K3$, but they still cancel in the kinetic term.}

The kinetic term mixing between the universal and nonuniversal axions
is \beq \int \t E_{5,r}\wedge\star_{10} \t E_{5,1}+\frac{g_s}{2}\int
E_{3,r}\wedge\star_{10} \bar E_{3,1}+\frac{g_s}{2}\int \bar
E_{3,r}\wedge\star_{10} E_{3,1} \eeq Using the constraint equations
(\ref{eq:manyconstraint},\ref{eq:g3constraint2}) in a calculation
similar to that presented below in section \ref{sec:4d}, the kinetic
mixing is proportional to  $(\t\rho_r,\t J)=\int \t\star\rho_r\wedge\t
J$.  Since this is the natural inner product on the 2nd cohomology,
the universal axion is orthogonal to the other $h^{1,1}-1$
axionic excitations as long as the basis of $(1,1)$ forms is chosen to
be orthogonal itself.

\section{K\"ahler potential}\label{sec:4d}

Finally we are ready to compute the kinetic term and K\"ahler
potential for the chiral superfield
\begin{equation}
\rho=a_0+ i\,c
\end{equation}
which combines the universal K\"ahler modulus found in \eq{delta-c-g}
with the axionic mode given in \eq{axion2}. Finding an explicit answer
for  the K\"ahler potential is in general rather involved, because the
compensating fields  appear explicitly in the kinetic
terms. Therefore, one would have to solve the  second order constraint
equations (which depend on the warp factor) and then plug  in the
explicit solution into the kinetic terms. However,  using the
Hamiltonian expressions for the kinetic terms, we will find  that the
explicit solution to the compensating fields is actually not
needed. We show that the constraint equations are enough to eliminate
the compensating fields from the  4d action. In this way, we compute
the explicit K\"ahler potential.

\subsection{Kinetic terms}

First we look at the kinetic term for $c(x)$,
\begin{equation}\label{eq:Kinc}
S_{kin,\,c}=\frac{1}{\kappa_{4}^2}\,\int d^4 x\,\sqrt{\hat
g}\,G_{cc}\,\hat g^{\mu\nu}\,\partial_\mu c\,\partial_\nu c\,.
\end{equation}
According to section \ref{sec:review}, $G_{cc}$ follows from replacing
the  canonical momentum conjugate to (\ref{eq:delta-c-g}) in the
Hamiltonian  expression (\ref{eq:hamilton-G}). A short computation
reveals that
\begin{equation}\label{eq:Gcc1}
G_{cc}= \frac{1}{2\,V_{CY}}\,\int d^6y \,\sqrt{\tilde
g}\,e^{4\Omega+2A}\,\left[e^{-2A-2\Omega}(\partial_c \Omega+\partial_c
A)-e^{2A} (\partial^{\tilde{m}} A)(\partial_m B)\right]\,.
\end{equation}

Integrating by parts to get  $\tilde \Del^2 B$ and replacing it by its
constraint (\ref{eq:constrB}), the terms containing $\partial_c A$
cancel, and $\partial_c \Omega$ controls the kinetic term. The result
is
\begin{equation}\label{eq:Gcc2}
G_{cc}=\frac{3}{4}\,e^{4\Omega} =\frac{3}{4}\,\left(\frac{V_{CY}}{c(x)
V_{CY} + V_W^0}\right)^2\,,
\end{equation}
showing the well-known factor of $3$ for the kinetic term of the
universal volume modulus.  It is interesting that this factor arises
from nontrivial cancellations of different warping corrections,  which
would not occur had we neglected the compensating field contribution.

To calculate the kinetic term for the universal axion, we take the
prescription for the  $5$-form in which we double the coefficient of
the $\t F_5^2$ term in the action but consider  only half the
components.  We will keep the terms including the scalar $a_0$ as
opposed to $a_2$ (with $a_0=a_0(t)$, this corresponds to keeping
components of $\t F_5$ with  time indices). Replacing the axion
fluctuation (\ref{eq:axion2}) into the kinetic action,  we
find\footnote{Recall that $E_p$ is the ``electric field''
$F_{0i_1...i_{p-1}}$.}  \barray
S_{kin,\,a}&=&-\frac{1}{4\kappa_{10}^2} \int \left(\t E_5\wedge \star_{10}
\t E_5 + g_s\, E_3\wedge \star_{10} \bar E_3\right)\nonumber\\ &=&
-\frac{1}{4\kappa_{10}^2}\int e^{4\Omega}da_0\wedge\hat\star_4 da_0
\int \left[\left(\t\star_6(e^{-4A}\t J + 4 dA\wedge dK) \right)\wedge
\left(\t J+de^{4A}\wedge dK\right) \right.\nonumber\\
&&\left.+e^{-2\Omega}\,g_s\,d\Lambda_1\wedge\t\star_6 d\bar\Lambda_1
\vphantom{\left(\frac{ig_s}{2}\right)}\right]\, .
\label{eq:axionkinetic1}
\earray Note that the Chern-Simons term does not include $a_0$, so it
does not appear.  Integrating by parts and using the constraint
equations  (\ref{eq:c-ansatz},\ref{eq:g3constraint2}) to eliminate the
compensators $K(y),\Lambda_1(y)$, we arrive to
\begin{equation}\label{eq:axionkinetic2}
S_{kin,\,a}=-\frac{3}{4\kappa_{4}^2} \int \sqrt{-\hat g} e^{4\Omega}
\hat g^{\mu\nu}\del_\mu a_0 \del_\nu a_0 \,.
\end{equation}
The factor of $3$ comes from
\begin{equation}\label{eq:j3}
\int \t J\,\wedge\,\t\star_6\t J = \frac{1}{2}\int \t J^3=3V_{CY}\,.
\end{equation}
This reproduces precisely the field space metric of the volume
modulus.  As we saw with the metric volume modulus, we see that the
presence of the compensators in (\ref{eq:axionkinetic1}) are crucial
to obtain the correct form for the kinetic term
(\ref{eq:axionkinetic2}).

\subsection{K\"ahler potential and no-scale structure}

The previous analysis shows that the volume modulus and universal
axion can be complexified into
\begin{equation}\label{eq:rho}
\rho(x)=a_0(x)+ i\,c(x)\,.
\end{equation}
In fact, since our analysis has not relied on the particular
components of the 3-form flux, the volume modulus and axion form a
complex scalar even in compactifications with classically broken
supersymmetry.  From the kinetic terms (\ref{eq:Kinc}) and
(\ref{eq:axionkinetic2}), we obtain
\begin{equation}
S_{kin} =-3 \frac{1}{\kappa_{4}^2}\int d^4x \sqrt{\hat g_4}\;
\frac{\hat g^{\mu\nu}\,\partial_\mu \rho \partial_\nu
\bar{\rho}}{\left[-i( \rho-\bar{\rho})+2\,V_W^0/V_{CY} \right]^2}
\end{equation}
This metric follows from the K\"ahler potential,
\begin{equation}
K = -3\log \left(-i(\rho-\bar{\rho}) +2\,\frac{V_W^0}{V_{CY}}\right)
\, .
\label{eq:UniversalKahlerPotential}
\end{equation}
Corrections due to warping amount to an additive constant in the
K\"ahler potential. This proves that no-scale structure
$G^{\rho\bar{\rho}} \partial_\rho K \partial_{\bar \rho} K =3$ is
maintained in GKP type compactifications, albeit in terms of a highly
nontrivial 10d wavefunction for $\rho$.  We can also write this 
K\"ahler potential in a more physical way in terms of the full
warped volume,
\begin{equation}
K = -3 \log \left(2 V_W(\rho)/V_{CY}\right)\,.
\end{equation}
\comment{
Although this expression for the K\"ahler potential looks very 
similar to those written in \cite{dg}, we note that the 
corresponding $10$-dimensional fluctuations are in fact quite
different, as we saw in previous sections.
}

The quantity $(V_W^0/V_{CY})$ may be interpreted as the background
value  for $c(x)$, so, after shifting \beq \rho \to
\rho-i\,\frac{V_W^0}{V_{CY}}\,, \eeq the K\"ahler potential is
\begin{equation}
K = -3\log \left[-i(\rho-\bar{\rho})\right] \,.
\label{eq:UniversalKahlerPotential2}
\end{equation}
This result coincides with the unwarped expression.  The correction
from warping becomes  important, for example, once a nonperturbative
superpotential for  $\rho$ is included as in \cite{kklt}. The
instanton or gaugino condensation superpotential receives then an
exponential  correction from warping due to the tree-level 
shift,\footnote{Here we are ignoring possible corrections to the K\"ahler
potential in the $\alpha'$ and $g_s$ expansions, as well as nonperturbative
corrections.}
\beq
W=A\,e^{ia\,\rho} \to A\,e^{aV_W^0/V_{CY}}\,e^{i a \rho}\,.  \eeq
Similarly, if we consider $\alpha'$ corrections \cite{hep-th/0204254},
the shift modifies the potential for the volume modulus.  The
modifications in both these cases deserve further study.


The fact that a series of rather subtle corrections conspire to give
the very  simple final result (\ref{eq:UniversalKahlerPotential})
suggests that there  could be some underlying physical reason for
this.\footnote{We thank S. Kachru  and A. Tomasiello for discussions on
this point.} One way to understand this  is to notice that (in the
absence of contributions beyond classical supergravity) the 10d
solution we have found preserves the shift symmetry  $e^{-4A} \to
e^{-4A} \,+ \,c(x)$.
This implies no-scale structure, which in turn restricts  the K\"ahler
potential to be of the general form \beq K(\rho, \bar \rho)=-3
\log\left[-i(\rho-\bar \rho)+a \right]+b\,.  \eeq Therefore, the
shift-symmetry of the full solution protects the  K\"ahler potential
from significant warping corrections.

\section{Nonlinear solution for fluctuating volume modulus}
\label{sec:ppwave}

In this section, we present a complete, nonlinear solution to the 10d
supergravity field equations corresponding to a wave of the  universal
volume modulus.  Our solutions are appropriate for compactifications
of the form discussed in \cite{gkp}.  For ease of  presenation, we
will work with the covariant equations of motion.

The external spacetime metric in the time-dependent background takes a
pp-wave form, as is appropriate for a propagating massless field.  As
a brief review, the pp-wave metric has the form
\beq\label{eq:ppmetric} ds^2_4 = \hat g_{\mu\nu}dx^\mu dx^\nu =
-H(x^+,\vec x)(dx^+)^2+dx^+dx^- +d\vec x^2\ .  \eeq A clear but
important property of this metric is that $\hat g^{++}=0$.  It will be
important later that most of the Christoffel symbols vanish; in
particular, $\hat\Gamma^+_{\mu\nu}=0$.  The only nonvanishing Ricci
tensor component is  $\hat
R_{++}=(1/2)\vec\del^{\,^{\textnormal{\scriptsize 2}}} H$,  so the
Ricci scalar vanishes.

As a source, consider a massless scalar with action
\beq\label{eq:genscalar} S=-\frac{1}{2\kappa_D^2}\int d^Dx \sqrt{-\hat
g} f(\phi) \left(\del\phi\right)^{\hat 2}\ .\eeq It is clear that any
function $\phi(x^+)$ solves the scalar equation of motion, and, since
$\del\phi$ is null, the Einstein equation is (the only nontrivial
component is $++$) \beq\label{eq:ppeinstein} \hat
R_{\mu\nu}=f(\phi)\del_\mu\phi \del_\nu\phi\ ,\eeq which is solved by
\beq\label{eq:ppsolution} H(x^+,\vec x)=\frac{1}{2(D-2)}\left|\vec
x\right|^2 f\left(\phi(x^+)\right) \left(\del_+\phi(x^+)\right)^2\
.\eeq Since $H$ is quadratic in the scalar velocity, we see
immediately why previous attempts to solve for the volume modulus
beyond linear order have failed.

\subsection{Ten-dimensional solution}

We can now present the nonlinear solution for a propagating volume
modulus and verify that it solves the equations of motion.  The warp
factor profile in the compact dimensions remains the same as in the
static case, and the compensator wavefunction is given by the
linearized expression.  In addition, since 3-form fluxes do not
stabilize the volume modulus, we include the 3-forms quite simply, so
these results apply to all GKP compactifications \cite{gkp}.
Throughout, we assume that 7-branes are in the orientifold limit, so
that the internal space is conformally CY and the axio-dilaton is
constant.  We also work away from localized sources such as branes or
orientifolds for simplicity; removing these assumptions is a
straightforward  generalization.

The 10d background corresponding to a finite fluctuation of the
universal volume modulus can be written as \barray ds^2&=&
e^{2A(x,y)}e^{2\Omega(x)}\bar g_{\mu\nu}(x,y)dx^\mu dx^\nu
+e^{-2A(x,y)} \t g_{ij}(y) dy^i dy^j\label{eq:fullg}\\ \t
F_5&=&e^{4\Omega} d^4x\wedge d\left(e^{4A}\right) + \t\star
d\left(e^{-4A_0}\right)\ ,\label{eq:full5} \earray where we have
defined the shorthand $e^{2\Omega}$ for the Einstein frame factor as
in \eq{Omega-def} and the warp factor as in \eq{A-c} as well as a 4d
metric \beq \bar g_{\mu\nu}(x,y)= \hat g_{\mu\nu}(x)
-2\left(\hat\Del_\mu \del_\nu c(x) +e^{2\Omega(x)}\del_\mu
c(x)\del_\nu c(x)\right) B(y)\label{eq:gbar}\ .  \eeq Here, $\hat
g_{\mu\nu}$ is a pp wave as defined in \eq{ppmetric}, and $B(y)$ is a
compensator that obeys the same constraint as in the linear case
\eq{constr4d}.  In addition, the volume modulus $c(x)$ depends only on
a null direction, which we denote $x^+$.  This means that
$\hat\Del_\mu \del_\nu c =\bar\Del_\mu \del_\nu c=\del_+^2 c$ (or for
any field). In addition, since $\hat g_{\mu\nu}$ and $\bar g_{\mu\nu}$
differ from Minkowski only in the $++$ component, $d^4x$ is the volume
form for those metrics as well (conveniently written in light-cone
coordinates).

The first equation of motion to check is the 5-form Bianchi identity,
which is satisfied as long as $A_0$ is the appropriate static warp
factor; with fixed background 3-form flux (and local sources), the
Bianchi identity is spacetime independent.  Self-duality of the 5-form
then fixes the spacetime component ---  the external component of
$C_4$ is just the volume form of the 4d spacetime.  It is also easy to
see that the axio-dilaton and 3-form equations of motion are unchanged
from the static solution (up to overall factors), so they are
trivially satisfied, as well.

We now proceed to the Einstein equation.  The $\mu i$ component is
just the integrated form of \eq{shift}, which is satisfied by the
``shifted'' form (\ref{eq:A-c}) assumed.  The internal component is
slightly more complicated because it includes sources from the 5-form
and 3-forms.  However,  because all 4d derivatives are null and the pp
wave Ricci scalar vanishes, the Einstein equation reduces to the
static case, which is satisfied by assumption. This is the Poisson
equation \beq\label{eq:internal} \t\Del^2 e^{-4A_0} = -\frac{g_s}{12}
G_{ijk}\bar G^{\widetilde{ijk}}\ ,\eeq which also follows from the
5-form Bianchi \cite{gm}.

Finally, we consider the external components of the Einstein equation.
A straightforward but somewhat tedious calculation finds the Ricci
tensor \barray R_{\mu\nu} &=& \hat R_{\mu\nu}-2\hat\Del_\mu\del_\nu
\Omega +4\hat\Del_\mu\del_\nu
A+2\del_\mu\Omega\del_\nu\Omega-8\del_{(\mu}\Omega \del_{\nu)}A
-16\del_\mu A\del_\nu A\nonumber\\ &&-e^{2\Omega}e^{4A}\bar
g_{\mu\nu}\t\Del^2 A+e^{2\Omega}e^{4A} \left(\hat\Del_\mu \del_\nu c
+e^{2\Omega}\del_\mu c\del_\nu c\right) B\ .\label{eq:fullricci}
\earray As in calculating the other components,  we have made repeated
use of the fact that all  spacetime derivatives lie in the $x^+$
direction, so contractions of them automatically vanish.   The
trace-reversed stress tensor (we take $R_{MN}=T_{MN}$) has external
components \beq T_{\mu\nu}= -4e^{2\Omega}e^{4A} \left(\del_i
A\del^{\t\imath}A\right) \bar
g_{\mu\nu}-\frac{g_s}{48}e^{2\Omega}e^{4A} G_{ijk}\bar
G^{\widetilde{ijk}} \bar g_{\mu\nu}\ .
\label{eq:fullstress}
\eeq Then the external Einstein equation simplifies with the help of
\eq{internal} along with the relations
(\ref{eq:Omega-def},\ref{eq:A-c}): \beq\label{eq:external} \hat
R_{\mu\nu}+\hat\Del_\mu\del_\nu c\left[e^{2\Omega}-e^{4A}+e^{2\Omega}
e^{4A}\t\Del^2 B\right]+\del_\mu c\del_\nu
c\left[-\frac{1}{2}e^{4\Omega}
-e^{2\Omega}e^{4A}+e^{4\Omega}e^{4A}\t\Del^2 B\right] = 0\ .  \eeq
Since we take the compensator $B$ to obey the constraint
(\ref{eq:constr4d}), we end up with \beq\label{eq:final4d} \hat
R_{\mu\nu}=\frac{3}{2}e^{4\Omega}\del_\mu c\del_\mu c\ .\eeq Note that
the compensator term quadratic in $c$  is necessary to cancel all the
internal space dependence in the external Einstein equation.  This is
just the Einstein equation (\ref{eq:ppeinstein}) for the 4d pp wave,
as we desired.

\subsection{Comments on the nonlinear background}

Let us now make a few comments about the nonlinear background.

First, compare this background to the linearized one presented
earlier.  The Hamiltonian approach naturally defines the compensators
as metric components $g_{\mu i}\propto \del_i B$.   These can be
gauged away at the cost of introducing a deformation of the internal
metric.  However, in the nonlinear solution, it is useful to work with
coordinates in which $\t g_{ij}$ is unchanged by the fluctuation and
the compensator appears in the  spacetime metric.  In addition, the
compensator now acquires a term  quadratic in the modulus
velocity. Finally, since the solution singles out the lightcone
coordinate $x^+$, we found it more convenient to work with the
covariant equations of motion.  Otherwise, the nonlinear background is
quite similar to the linearized one, and we see that the warp factor
and compensator profiles are actually identical.

The existence of this nonlinear background  has several important
consequences.  For one, the solution provides an independent
derivation of the kinetic term for the volume modulus.  That is, the
10d solution satisfies the 4d Einstein equation for the pp-wave
(\ref{eq:ppeinstein}), which  exactly encodes the kinetic term for the
massless scalar.   In fact, we see that we reproduce the field space
metric (\ref{eq:Gcc2}), even including the famous factor of 3.  This
fact is a highly nontrivial consistency check of the low energy theory
that we have developed.

This solution is also the first time-dependent 10d background that
correctly captures the nonlinear physics of modulus motion in warped
string compactifications.  Since it is precisely consistent with the
expected effective field theory, it should end concerns raised in
\cite{ku1,ku2}  about the validity of the 4d effective theory.

Finally, it seems that this solution is likely to share a number of
features with cosmological backgrounds in these compactifications; in
particular, if the K\"ahler modulus is  stabilized with a mass well
below the warped KK scale, its motion will  be well approximated by
classical solutions.  Developing cosmological backgrounds would be of
relevance to models of inflation in string theory and could shed light
on higher-dimensional or string physics in cosmology.  Unfortunately,
solving for the motion of the K\"ahler modulus in a cosmological
background is already difficult at the 4d level, so we leave this
issue as an open question.

\section{Strongly warped limit and light KK modes}\label{sec:strongwavefcn}

In the previous sections we have obtained the 10d solution
corresponding to the universal K\"ahler modulus, first in the
linearized approximation, and then showing how to include finite
fluctuations. We also studied the 4d properties of the solution, by
finding the K\"ahler potential and proving no-scale structure. In this
section we will show how to apply our results to strongly warped
throats in the compactification manifold.

Strongly warped regions are important both from a phenomenological
point of view and to understand gauge/gravity dualities in string
theory. Moreover, the effects from compensating fields are expected to
dominate in this limit \cite{dt}, so this is good place to illustrate
our results.  Another important dynamical effect is that at strong
warping the KK mass scale is redshifted, and could become of the same
order as the energy scale of the EFT for the moduli fields. Therefore,
these new light fields need to be included in the 4d description. In
the first part of the section we will find the 10d wavefunction of the
volume modulus at strong warping, and illustrate its behavior for
various choices of warping. Next we show will how to include light KK
modes, concluding that there are no kinetic mixings with the K\"ahler
modulus.

\subsection{Wavefunction in the strongly warped limit}

To begin with a simple example, consider an AdS warp factor
$e^{-4A_0}\sim N/r^4$. Without including compensating fields, the 10d
wavefunction corresponding to the volume modulus  $c(x)$ scales, at
small $r$, like
\begin{equation}
\delta_c g_{\mu\nu}\, \sim \,\frac{r^6}{N^{3/2}}\;,\;\; \delta_c
g_{rr} \,\sim \,\frac{r^2}{N^{1/2}}\,.
\end{equation}
On the other hand, including the effect of compensating fields,  we
obtain the qualitatively different behavior
\begin{equation}
\label{eq:AdSWaveFcn}
\delta_c g_{\mu\nu} \,\sim \, \frac{r^2}{N^{1/2}} \;,\;\; \delta_c
g_{rr} \,\sim \, \frac{N^{1/2}}{r^2}\,.
\end{equation}
This illustrates the point that the correct gauge invariant 10d
fluctuation may differ significantly from the naive solution.

Let us be more concrete and model the throat locally by a warped
deformed conifold with metric given by the the Klebanov-Strassler
solution \cite{ks},
\begin{eqnarray}
ds^2 &=& e^{2A_0} \eta_{\mu\nu} + e^{-2 A_0} \frac{\epsilon^{4/3}}{2}
K(\tau)  \Big[ \frac{d\tau^2 + (g^5)^2}{3 K^3(\tau)} \nonumber \\ && +
\cosh^2\left(\frac{\tau}{2}\right) ((g^3)^2+(g^4)^2) +
\sinh^2\left(\frac{\tau}{2}\right) ((g^1)^2+(g^2)^2)\Big]
\end{eqnarray}
where $\tau$ is the radial coordinate along the throat.  The equation
for the compensator (\ref{eq:constr4d}) now becomes
\begin{equation}
\partial_\tau \left(K^2(\tau) \cosh^2\frac{\tau}{2}\
\sinh^2\frac{\tau}{2}\  B_\tau(\tau)\right) =
\left(\frac{V_W^0}{V_{CY}}-e^{-4A_0(y)}\right)
\frac{\epsilon^{4/3}}{6}  \cosh^2\frac{\tau}{2}\ \sinh^2\frac{\tau}{2}
\end{equation}

\FIGURE{ \includegraphics[scale=.43]{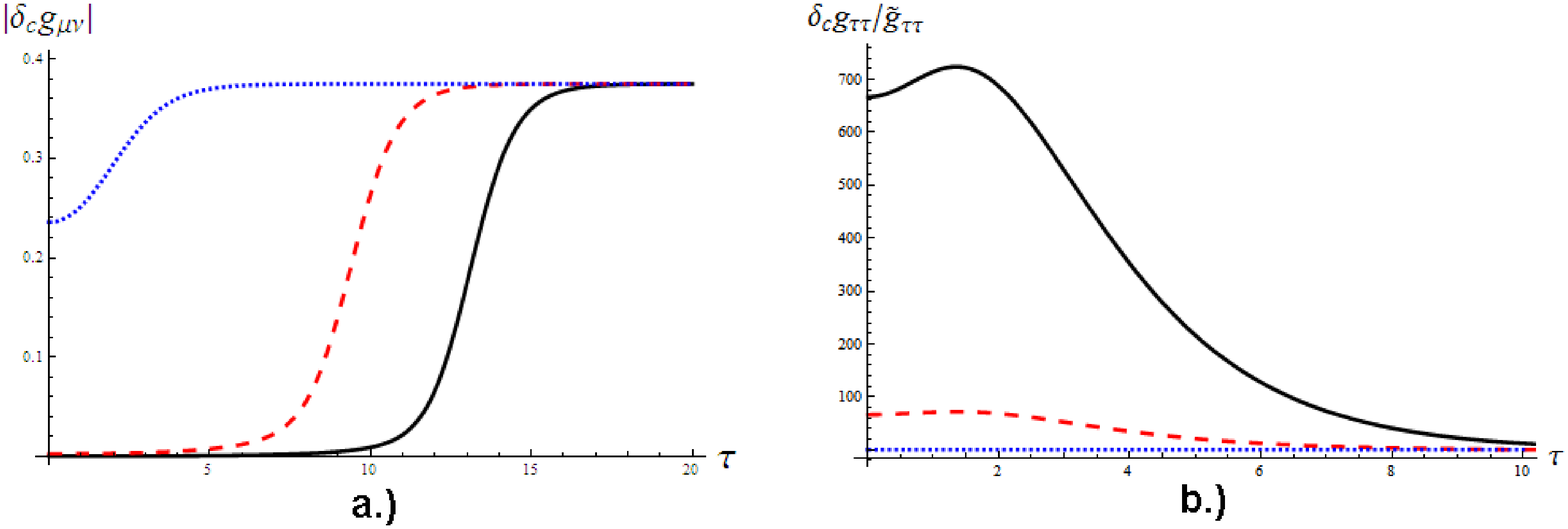}
\caption{(a) The $4$-dimensional wavefunction
$\delta_c g_{\mu\nu}$ and (b) the internal metric wavefunction
$\delta_c g_{\tau \tau}/\tilde{g}_{\tau\tau}$ in a Klebanov-Strassler
warped background for various values of the warping evaluated at the
tip $e^{-4A_0(0)}$: no warping $e^{-4A_0(0)} = 1$, dotted blue; weak
warping $e^{-4A_0(0)} = 10^4$, dashed red; strong warping
$e^{-4A_0(0)} = 10^6$, solid black.  Notice that as the warping
increases, the wavefunction dips deeper into the throat.}
\label{fig:WarpedWaveFcns}
}

One can now solve this equation numerically for various values of  the
warping -- the results for the wavefunctions  $\delta_c
g_{\mu\nu},\delta_c g_{\tau \tau}/\tilde{g}_{\tau\tau}$ are shown in
Figure \ref{fig:WarpedWaveFcns}. For convenience of display in Figure
\ref{fig:WarpedWaveFcns} we have divided out the unwarped part
$\tilde{g}_{\tau\tau}$ of the metric to show that at large $\tau$,
where the warping is weak, the physical metric flucutation asymptotes
to the unfluctuated and unwarped metric, which is what we expect.

As the amount of warping increases (dashed red and solid black lines)
the internal metric wavefunctions $\delta_c g_{ij}$ become  more
peaked in the tip region of the throat where the warping is strongest,
while the 4d metric wavefunctions $\delta_c g_{\mu\nu}$ decrease to
zero, as expected from our simple estimates with the AdS warp factor
(\ref{eq:AdSWaveFcn}).

\subsection{Inclusion of KK modes}

We now address the problem of including light KK modes in the EFT of
the volume modulus.\footnote{We thank E. Silverstein for suggesting to
check this.}  A general argument for the absence of kinetic mixings
beween zero modes and their KK excitations was given in
\cite{stud}. It was based on the observation that these fluctuations
are eigenvectors of a Sturm-Liouville problem, such that the
orthogonality relation derived from the differential problem coincides
with the Hamiltonian inner product. This then grants the absence of
kinetic mixings. Since the application to $p$-forms may be unfamiliar,
we now show that the universal axion is orthogonal to its KK
excitations.

Consider then the 2-form massless and massive modes in $C_4$,
\begin{equation}
\delta C_4= a_2(x) \wedge \tilde J(y) + 
\sum_\alpha a_2^\alpha(x) \wedge \omega_\alpha(y)
\end{equation}
where $\omega_\alpha$ are (non-closed) 2-forms, and the 
KK fields $a_2^\alpha$ are dual to spacetime scalars. The compensating 
fields are already absorbed into $\tilde J$ and $\omega_\alpha$. 
For simplicity, we are also setting the Weyl factor equal to one. 
There are, of course, other components, and we have not determined the 
complete wavefunctions for the excited KK modes, but we can see orthogonality
just from these components.

Requiring that the particles have a well-defined 4d mass, 
$d\left( \hat \star_4 da_2^\alpha \right)=-m^2_{\alpha}\,
\hat \star_4 a_2^\alpha$, we derive the eigenvector equation
\begin{equation}\label{eq:axion-eigenv}
d\left(\tilde \star_6 d \omega_\alpha \right)=m_\alpha^2\,e^{-4A}\,
\tilde \star_6 \omega_\alpha\,.
\end{equation}
The computation of the kinetic mixing between $a_2(x)$ and $a_2^\alpha(x)$ 
then proceeds as in \eq{axionkinetic1}:
\begin{eqnarray}
\int E_5 \wedge\, \star_{10}\, E_5 &\to& - \int_x a_2(x) \wedge\,d 
\left[\hat \star_4\,da_2^\alpha(x) \right]\,
\int_y\,e^{-4A(y)}\,\tilde J \wedge\,\tilde \star_6 \omega_\alpha\nonumber\\
&=&-\frac{1}{m_\alpha^2}\,\int_x a_2(x) \wedge\,d \left[\hat \star_4\,
da_2^\alpha(x) \right]\,\int_y\,\tilde J \wedge 
d\left(\tilde \star_6 d \omega_\alpha \right)
\end{eqnarray}
where we have used (\ref{eq:axion-eigenv}). 
Since $\tilde J$ is closed, integrating by parts the kinetic mixing vanishes.

By supersymmetry, the same holds for the universal volume modulus
(since the analysis should not depend on our choice of 3-form flux,
this statement holds even in classically nonsupersymmetric compactifications). 
We conclude that light KK modes do not mix with the K\"ahler modulus 
at the level of the kinetic terms.

\section{Discussion and implications}
\label{sec:discussion}

By using the Hamiltonian method, developed for warped compactifications
in \cite{dt}, we have computed the kinetic term and K\"ahler potential 
for the universal volume modulus and its axionic partner
in IIB flux compactifications of the type studied in \cite{gkp}
for arbitrary warping.
We found that the K\"ahler potential for the universal
K\"ahler modulus takes the form
\begin{equation}
K(\rho, \bar \rho) = -3 \log\left(-i(\rho-\bar{\rho})-2\,
\frac{V^0_W}{V_{CY}}\right)\,.
\end{equation}
It is rather striking that all warping corrections just 
amount to an additive shift 
$\rho \rightarrow \rho-i\,(V_W/V_{CY})$. One way to understand 
this result is to 
argue that the no-scale symmetry survives in the correct 10d warped 
solution. This 
protects the K\"ahler potential from further warping corrections.

It is important to emphasize that the 10d time-dependent solution that 
we have found 
is very different from the unwarped fluctuation. Therefore, the 
respective 4d theories 
are expected to be different as well, even if the K\"ahler potentials 
have the same 
functional dependence. In particular, once nonperturbative corrections 
of the form 
$W=A\,e^{i a \rho}$ are included, the previous seemingly 
innocuous shift in $\rho$ 
may produce qualitative changes in the field theory. This could 
become important 
in KKLT type models \cite{kklt} that rely 
on the existence of a strongly warped region. 
It would be interesting to compute the prefactor $A$ 
(see \cite{Ganor:1996pe,Berg:2004ek,Witten:1996bn}) 
in strongly warped backgrounds, 
and see how our 10d solution modifies the discussion.

In section \ref{sec:strongwavefcn} we showed that the warped 10d
fluctuations for a time-dependent universal volume modulus are peaked
at the tip of the throat, and that there are no K\"ahler  potential
mixings with light KK modes.   This can be relevant for
phenomenological applications in which the coupling of the universal
K\"ahler modulus to brane and bulk fields, obtained by the 10d
wavefunction overlap, is important.  Also, studying further the
wavefunctions of the KK modes of the  universal axion could shed light
on the possibility of mixing through  mass terms as well as be
important for studying the behavior of perturbations in strongly
warped throats.

We have also shown in section \ref{sec:ppwave} that the 10d metric
fluctuations can be promoted to a fully time-dependent, warped, 10d
metric for the universal volume modulus by taking into account the
backreaction on the 4d space.  This is a first step towards finding
cosmological solutions for time-dependent K\"ahler moduli, which may
be relevant for models of inflation.

There are several future directions of interest.  First, it is highly
desirable to determine the K\"ahler potential for general K\"ahler moduli,
which are not stabilized by 3-form flux on a generic CY.  Another 
interesting related open problem is calculating the K\"ahler potential
for modes that are stabilized by the 3-form flux; as discussed in
\cite{dt, hep-th/0308156,hep-th/0603233,stud}, the flux also modifies the 
10d wavefunction in this case.  On a slightly
different tack, it is natural to extend our results to excited KK modes
of the volume modulus and axion, along the lines of \cite{stud}.
Finally, generalization of our nonlinear solution to cosmological 
backgrounds is an important problem for future work in string cosmology.

\acknowledgments
We would like to thank C. Burgess, 
F. Denef, S. Franco, T. Grimm, S. Kachru, A. Maharana,
A. Nacif, J. Polchinski, G. Shiu, E. Silverstein and A. Tomasiello for
useful discussions and comments. M.R.D. and G.T. are 
supported in part by DOE grant DE-FG02-96ER40959. G.T. would like to thank KITP for their hospitality and support through the National Science Foundation Grant PHY05-51164.
The work of A. R. F. is supported by the NSERC
Discovery Grant program.  
B. U. is supported by NSERC and by
fellowships from the Institute  for Particle Physics (Canada) and
Lorne Trottier (McGill).

\appendix
\section{Gauge transformations and field redefinitions of $C_4$}
\label{sec:5formdef}

The dimensional reduction of fluctuations of $C_4$ in 3-form flux 
background is slightly subtle due to its nonstandard gauge transformations.
We follow the discussion of \cite{hep-th/0201029}, which considered
the case of a torus orientifold in some detail.

In terms of the 4-form that couples electrically to a D3-brane
\beq\label{eq:wz}
S_{WZ}=\mu_3\int C_4\ ,\eeq
the 5-form field strength is $\t F_5=dC_4-C_2 H_3$.
The gauge transformations that leave $\t F_5$ invariant are
\beq\label{eq:gauge}
C_4\to C_4+d\chi_3 +\zeta^C_1 \wedge H_3\ , \ \ C_2\to C_2+d\zeta^C_1\ ,\ \
B_2\to B_2+d\zeta^B_1\ .\eeq
In a background of nontrivial 3-form flux, 
the potentials $B_2$ and $C_2$ are well-defined
only on coordinate patches, which must be glued together with
gauge identifications $\zeta^{B,C}$.  With a fixed choice of background
potentials $C_4$, $B_2$, and $C_2$, the gauge transformations $\zeta^{B,C}$
are also fixed, so fluctuations $\delta B_2$, $\delta C_2$ must be 
globally defined on the internal manifold (on a torus, this means they are 
periodic).  Hence, they have the appropriate behavior for dimensional 
reduction without any issue of gluing coordinate patches together.

The 4-form is slightly more complicated; the background $C_4$ is
also defined only on patches and glued together by the gauge transformation
(\ref{eq:gauge}) with $H_3$ the background flux.  This means that
the fluctuation also has a nontrivial gauge gluing 
$\delta C_4\to \delta C_4+d\chi+\zeta^C \delta H_3$.  To simplify the 
gluing conditions, we can define $\delta C'_4=\delta C_4-C_2\delta B_2$
(to linear order); 
this is glued together by gauge transformations 
$\delta C'_4\to \delta C'_4 +d\chi'$ with
$\chi'=\chi-\lambda^C\delta B_2$, which are trivial as long as there
is no quantized 5-form flux.  Therefore, the 4-form potential that follows
ordinary dimensional reduction is $\delta C'_4$.  
The field strength and
complete gauge transformations work out to be
\barray
\delta \t F_5 &=& d\delta C'_4+\frac{ig_s}{2}\left(\delta A_2\wedge \bar G_3
-\delta\bar A_2\wedge G_3\right)\label{eq:f5def}\\
\delta C'_4&\to& \delta C'_4+\delta\chi'+\frac{ig_s}{2}\left(\bar\zeta^A
\wedge G_3-\zeta^A\wedge \bar G_3\right)\label{eq:primegauge}
\earray
in terms of the complex potenial $A_2=C_2-\tau B_2$, $G_3=dA_2$.
Henceforth, we drop the prime on $\delta C_4$.

Lest this seem like a technical but nonphysical point, let us make two 
comments.  First, this field redefinition allows us to define the 
fluctuation in the 5-form without reference to the background 2-form
potentials, which is an immense simplification.  Second, the redefined
4-form fluctuation does not couple directly to the D3-brane as in
\eq{wz}.  The field redefinition modifies the coupling of the 2-form 
fluctuations to the D3-branes.

\bibliographystyle{utcaps2}
\bibliography{ukahler}

\providecommand{\href}[2]{\texttt{#2}}\begingroup\raggedright\begin{thebibliog%
raphy}{10}

\bibitem{dt}
M.~R. Douglas and G.~Torroba, ``{Kinetic terms in warped compactifications},''
\href{http://www.arXiv.org/abs/0805.3700}{ 0805.3700}.

\bibitem{stud}
G.~Shiu, G.~Torroba, B.~Underwood, and M.~R. Douglas, ``{Dynamics of warped
  flux compactifications},'' {\em JHEP} {\bf 06} (2008) 024,
\href{http://www.arXiv.org/abs/0803.3068}{ 0803.3068}.

\bibitem{dst}
M.~R. Douglas, J.~Shelton, and G.~Torroba, ``{Warping and supersymmetry
  breaking},''
\href{http://www.arXiv.org/abs/0704.4001}{ 0704.4001}.

\bibitem{hep-th/9605053}
K.~Becker and M.~Becker, ``{M-Theory on Eight-Manifolds},'' {\em Nucl. Phys.}
  {\bf B477} (1996) 155--167,
\href{http://www.arXiv.org/abs/hep-th/9605053}{ hep-th/9605053}.

\bibitem{hep-th/9908088}
K.~Dasgupta, G.~Rajesh, and S.~Sethi, ``{M theory, orientifolds and G-flux},''
  {\em JHEP} {\bf 08} (1999) 023,
\href{http://www.arXiv.org/abs/hep-th/9908088}{ hep-th/9908088}.

\bibitem{hep-th/0004103}
B.~R. Greene, K.~Schalm, and G.~Shiu, ``{Warped compactifications in M and F
  theory},'' {\em Nucl. Phys.} {\bf B584} (2000) 480--508,
\href{http://www.arXiv.org/abs/hep-th/0004103}{ hep-th/0004103}.

\bibitem{gkp}
S.~B. Giddings, S.~Kachru, and J.~Polchinski, ``{Hierarchies from fluxes in
  string compactifications},'' {\em Phys. Rev.} {\bf D66} (2002) 106006,
\href{http://www.arXiv.org/abs/hep-th/0105097}{ hep-th/0105097}.

\bibitem{Candelas:1990pi}
P.~Candelas and X.~de~la Ossa, ``{Moduli space of Calabi-Yau manifolds},'' {\em
  Nucl. Phys.} {\bf B355} (1991)
455--481.

\bibitem{hep-th/0201029}
A.~R. Frey and J.~Polchinski, ``{$\N = 3$ warped compactifications},'' {\em
  Phys. Rev.} {\bf D65} (2002) 126009,
\href{http://www.arXiv.org/abs/hep-th/0201029}{ hep-th/0201029}.

\bibitem{hep-th/0308156}
A.~R. Frey, ``{Warped strings: Self-dual flux and contemporary
  compactifications},''
\href{http://www.arXiv.org/abs/hep-th/0308156}{ hep-th/0308156}.

\bibitem{da1}
S.~P. de~Alwis, ``{On potentials from fluxes},'' {\em Phys. Rev.} {\bf D68}
  (2003) 126001,
\href{http://www.arXiv.org/abs/hep-th/0307084}{ hep-th/0307084}.

\bibitem{buchel}
A.~Buchel, ``{On effective action of string theory flux compactifications},''
  {\em Phys. Rev.} {\bf D69} (2004) 106004,
\href{http://www.arXiv.org/abs/hep-th/0312076}{ hep-th/0312076}.

\bibitem{da2}
S.~P. de~Alwis, ``{Brane worlds in 5D and warped compactifications in IIB},''
  {\em Phys. Lett.} {\bf B603} (2004) 230--238,
\href{http://www.arXiv.org/abs/hep-th/0407126}{ hep-th/0407126}.

\bibitem{gm}
S.~B. Giddings and A.~Maharana, ``{Dynamics of warped compactifications and the
  shape of the warped landscape},'' {\em Phys. Rev.} {\bf D73} (2006) 126003,
\href{http://www.arXiv.org/abs/hep-th/0507158}{ hep-th/0507158}.

\bibitem{hep-th/0312232}
M.~Gra\~na, T.~W. Grimm, H.~Jockers, and J.~Louis, ``{Soft supersymmetry
  breaking in Calabi-Yau orientifolds with D-branes and fluxes},'' {\em Nucl.
  Phys.} {\bf B690} (2004) 21--61,
\href{http://www.arXiv.org/abs/hep-th/0312232}{ hep-th/0312232}.

\bibitem{hep-th/0409098}
H.~Jockers and J.~Louis, ``{The effective action of D7-branes in N = 1
  Calabi-Yau orientifolds},'' {\em Nucl. Phys.} {\bf B705} (2005) 167--211,
\href{http://www.arXiv.org/abs/hep-th/0409098}{ hep-th/0409098}.

\bibitem{hep-th/0507257}
L.~Kofman and P.~Yi, ``{Reheating the universe after string theory
  inflation},'' {\em Phys. Rev.} {\bf D72} (2005) 106001,
\href{http://www.arXiv.org/abs/hep-th/0507257}{ hep-th/0507257}.

\bibitem{hep-th/0602136}
X.~Chen and S.~H.~H. Tye, ``{Heating in brane inflation and hidden dark
  matter},'' {\em JCAP} {\bf 0606} (2006) 011,
\href{http://www.arXiv.org/abs/hep-th/0602136}{ hep-th/0602136}.

\bibitem{arxiv:0710.1299}
A.~Berndsen, J.~M. Cline, and H.~Stoica, ``{Kaluza-Klein relics from warped
  reheating},'' {\em Phys. Rev.} {\bf D77} (2008) 123522,
\href{http://www.arXiv.org/abs/0710.1299}{ 0710.1299}.

\bibitem{Harling:2008px}
B.~v. Harling and A.~Hebecker, ``{Sequestered Dark Matter},'' {\em JHEP} {\bf
  05} (2008) 031,
\href{http://www.arXiv.org/abs/0801.4015}{ 0801.4015}.

\bibitem{arxiv:0802.2958}
J.~F. Dufaux, L.~Kofman, and M.~Peloso, ``{Dangerous angular KK/glueball relics
  in string theory cosmology},''
\href{http://www.arXiv.org/abs/0802.2958}{ 0802.2958}.

\bibitem{Burgess:2006mn}
C.~P. Burgess {\em et al.}, ``{Warped supersymmetry breaking},'' {\em JHEP}
  {\bf 04} (2008) 053,
\href{http://www.arXiv.org/abs/hep-th/0610255}{ hep-th/0610255}.

\bibitem{ku1}
H.~Kodama and K.~Uzawa, ``{Moduli instability in warped compactifications of
  the type IIB supergravity},'' {\em JHEP} {\bf 07} (2005) 061,
\href{http://www.arXiv.org/abs/hep-th/0504193}{ hep-th/0504193}.

\bibitem{ku2}
H.~Kodama and K.~Uzawa, ``{Comments on the four-dimensional effective theory
  for warped compactification},'' {\em JHEP} {\bf 03} (2006) 053,
\href{http://www.arXiv.org/abs/hep-th/0512104}{ hep-th/0512104}.

\bibitem{kklt}
S.~Kachru, R.~Kallosh, A.~Linde, and S.~P. Trivedi, ``{De Sitter vacua in
  string theory},'' {\em Phys. Rev.} {\bf D68} (2003) 046005,
\href{http://www.arXiv.org/abs/hep-th/0301240}{ hep-th/0301240}.

\bibitem{hep-th/0204254}
K.~Becker, M.~Becker, M.~Haack, and J.~Louis, ``{Supersymmetry breaking and
  alpha'-corrections to flux induced potentials},'' {\em JHEP} {\bf 06} (2002)
  060,
\href{http://www.arXiv.org/abs/hep-th/0204254}{ hep-th/0204254}.

\bibitem{hep-th/0408054}
V.~Balasubramanian and P.~Berglund, ``{Stringy corrections to Kaehler
  potentials, SUSY breaking, and the cosmological constant problem},'' {\em
  JHEP} {\bf 11} (2004) 085,
\href{http://www.arXiv.org/abs/hep-th/0408054}{ hep-th/0408054}.

\bibitem{hep-th/0502058}
V.~Balasubramanian, P.~Berglund, J.~P. Conlon, and F.~Quevedo, ``{Systematics
  of moduli stabilisation in Calabi-Yau flux compactifications},'' {\em JHEP}
  {\bf 03} (2005) 007,
\href{http://www.arXiv.org/abs/hep-th/0502058}{ hep-th/0502058}.

\bibitem{hep-th/0508139}
A.~R. Frey, A.~Mazumdar, and R.~C. Myers, ``{Stringy effects during inflation
  and reheating},'' {\em Phys. Rev.} {\bf D73} (2006) 026003,
\href{http://www.arXiv.org/abs/hep-th/0508139}{ hep-th/0508139}.

\bibitem{dg}
O.~DeWolfe and S.~B. Giddings, ``{Scales and hierarchies in warped
  compactifications and brane worlds},'' {\em Phys. Rev.} {\bf D67} (2003)
  066008,
\href{http://www.arXiv.org/abs/hep-th/0208123}{ hep-th/0208123}.

\bibitem{gray}
J.~Gray and A.~Lukas, ``{Gauge five brane moduli in four-dimensional heterotic
  models},'' {\em Phys. Rev.} {\bf D70} (2004) 086003,
\href{http://www.arXiv.org/abs/hep-th/0309096}{ hep-th/0309096}.

\bibitem{adm}
R.~L. Arnowitt, S.~Deser, and C.~W. Misner, ``{Canonical variables for general
  relativity},'' {\em Phys. Rev.} {\bf 117} (1960)
1595--1602.

\bibitem{moore}
D.~Belov and G.~W. Moore, ``{Holographic action for the self-dual field},''
\href{http://www.arXiv.org/abs/hep-th/0605038}{ hep-th/0605038}.

\bibitem{ks}
I.~R. Klebanov and M.~J. Strassler, ``{Supergravity and a confining gauge
  theory: Duality cascades and chiSB-resolution of naked singularities},'' {\em
  JHEP} {\bf 08} (2000) 052,
\href{http://www.arXiv.org/abs/hep-th/0007191}{ hep-th/0007191}.

\bibitem{Ganor:1996pe}
O.~J. Ganor, ``{A note on zeroes of superpotentials in F-theory},'' {\em Nucl.
  Phys.} {\bf B499} (1997) 55--66,
\href{http://www.arXiv.org/abs/hep-th/9612077}{ hep-th/9612077}.

\bibitem{Berg:2004ek}
M.~Berg, M.~Haack, and B.~Kors, ``{Loop corrections to volume moduli and
  inflation in string theory},'' {\em Phys. Rev.} {\bf D71} (2005) 026005,
\href{http://www.arXiv.org/abs/hep-th/0404087}{ hep-th/0404087}.

\bibitem{Witten:1996bn}
E.~Witten, ``{Non-perturbative superpotentials in string theory},'' {\em Nucl.
  Phys.} {\bf B474} (1996) 343--360,
\href{http://www.arXiv.org/abs/hep-th/9604030}{ hep-th/9604030}.

\bibitem{hep-th/0603233}
A.~R. Frey and A.~Maharana, ``{Warped spectroscopy: Localization of frozen bulk
  modes},'' {\em JHEP} {\bf 08} (2006) 021,
\href{http://www.arXiv.org/abs/hep-th/0603233}{ hep-th/0603233}.

\end{thebibliography}\endgroup

\end{document}